\documentclass[a4paper,12pt]{JHEP3}
\usepackage{amsmath}
\preprint{SISSA 26/2005/EP}

\title{Quadratic $\alpha'$ corrections to T-duality}

\author{{Ghasem Exirifard}\\
SISSA/ISAS, Via Beirut 2-4, I-34013 Trieste, Italy\\INFN, Sezione di Trieste, Italy\\
 Email: \email{exir@sissa.it}
 }

\abstract{The quadratic $\alpha'$ corrections to the two-dimensional black hole and to its T-dual are calculated. These backgrounds are used to write the  covariant form of the $\alpha'^2$ corrections to the T-duality  for general time-dependent backgrounds of dilaton and diagonal metric in the bosonic string theory. }


\begin{document}

\section{Introduction}
It is known that a chain of dualities relates different perturbative string theories on flat space-time to each other. The explicit form of possible generalization of these dualities to an arbitrary background is known only at the level of SUGRA or gravity. In general it is not known how these dualities should be modified as one takes into account the sub-leading string corrections. 

T-duality\cite{firsttduality} is the first known duality of string theory. It  connects two points of the moduli space of string theory. In the LEEA this connection  means that a map exists from the  field contents of a given background to the field contents of the corresponding T-dual background. Truncating the LEEA of string theory to the leading term, the so-called SUGRA or gravity, gives tree level rules of this map \cite{buscher}. These tree level rules should be modified as one adds the sub leading , in particular the word-sheet, corrections. The linear $\alpha'$-corrections to these rules in bosonic and heterotic string theory have been computed in \cite{Tseylinduality,olsen,kaloper,Schiapa}. 

Discrete T-duality can be generalized to a continuous symmetry group known as $O(d,d)$ transformation\cite{oddoneloop}. The argument that the string tree level effective action is invariant under continuous duality transformations to all orders in $\alpha'$ is given in \cite{sen1} and  is generalized for heterotic string theory in \cite{sen2}.  Also \cite{oddtwoloop,odd} provide the explicit computation that  $O(d,d)$ invariance is preserved  at order of $\alpha'^2$ given a field redefinition and coordinate transformations. In \cite{exir} the explicit $\alpha'^2$ corrections  to the rules of discrete T-duality were computed  for time-dependent diagonal metric and dilaton in the bosonic string. However the rule for the direction in which the T-duality is  applied  was provided only to  linear order in $\alpha'$. In this note we are going to compute the explicit form of the $\alpha'^2$ corrections to this rule. The note is organized in the following way:

We first review the results of \cite{exir} and write explicitly the ambiguity of the covariant form of T-duality. Next we calculate  the linear and the quadratic $\alpha'$ corrections to the two-dimensional black hole and to its corresponding T-dual background. We use these $\alpha'^2$ corrected backgrounds to write uniquely the sought $\alpha'^2$ corrections.

In the third section we discuss  the behaviour of the $\alpha'^2$ corrected two-dimensional black hole. In the fourth section we summarize  and  discuss the results.
\section{The $\alpha'$ corrections to T-duality}
The LEEA action of the free string theory is provided by the vanishing of the $\beta$ function of the corresponding sigma model. The $\beta$ functions of the free bosonic string theory for backgrounds of dilaton ($\phi$) and metric in $d$ dimensional space-time at the quadratic order in $\alpha'$ read \cite{Jack1,Tsytlin1}
\begin{eqnarray}\label{g2loop}
\frac{1}{\alpha'}
\beta_{ij}&=&R_{ij}~+~2~\bigtriangledown_i\bigtriangledown_j~\phi~+~\frac{1}{2}~\alpha'~R_{iklm}R_j^{~~klm}\\
&&+\alpha'^2\left\{\frac{1}{8}\bigtriangledown_k R_{ilmn}\bigtriangledown^k R_{j}^{~~lmn}-\frac{1}{16}\bigtriangledown_iR_{klmn}\bigtriangledown_jR^{klmn}+\frac{1}{2}R_{klmn}~R_{i}^{~~mlp}~R_{j~~~p}^{~~kn}
\right.\nonumber\\
&&\qquad~~~~~\left.~-~\frac{3}{8}~R_{iklj}~R^{kmnp}~R^{l}_{~~mnp}~+~\frac{1}{32}~\bigtriangledown_j\bigtriangledown_i(~R_{klmn}~R^{klmn})\right\}~,\nonumber\\
\frac{1}{\alpha'}
\beta_\phi&=&~\frac{d-26}{6\alpha'}~-~\frac{1}{2}~\Box~\phi~+~\partial_k\phi~\partial^k \phi~+~\frac{1}{16}~\alpha'~R_{klmn}~R^{klmn}\label{betaphinoncritical}\\
&&~+~\alpha'^2~\left\{-~\frac{3}{16}~R^{kmnp}~R^{l}_{~~mnp}~\bigtriangledown_k\bigtriangledown_l\phi~+~\frac{1}{32}~R_{klmn}~R^{mnpq}~R_{pq}^{~~~kl}\right.\nonumber\\
&&\left.\qquad~-~\frac{1}{24}~R_{klmn}~R^{qnpl}~R_{q~~p}^{~m~~k}~+~\frac{1}{64}~\partial_i(R_{klmn}~R^{klmn})~\partial^i\phi\right\}~,\nonumber\\
&&i,\cdots,q~\in~\{0...25\}\nonumber~.
\end{eqnarray}
where the computation is done in the dimensional regularization method in the minimal substraction scheme. These equations provide the perturbative dynamics of the metric and the dilaton around the flat space-time. Other regularization methods and renormalization schemes give  different $\beta$ functions. The $\beta$ functions of  different schemes should be related to one another by  perturbative field redefinition, e.g.\footnote{It is the field redefinition ambiguity at the linear order in $\alpha'$ for  schemes  in which  the metric is a tensor of rank two.} 
\begin{eqnarray}
g_{\mu\nu}&\to&g_{\mu\nu}+\alpha'{\big{(}}\,c_1\,R_{\mu\nu}+\,c_2\,R\,g_{\mu\nu}+
\,c_3\,\nabla_\mu\phi\,\nabla_\nu\phi+\,c_4\, \Box\,\phi\,g_{\mu\nu}+\,c_5\nabla_\mu\nabla_\nu\phi\,\big{)}+O(\alpha'^2) \,,\nonumber
\end{eqnarray}
where $c_1,\cdots,c_5$ are arbitrary functions of the dilaton. If we had been interested in finding the exact solution of the truncated $\beta$ functions\footnote{For example \cite{sen_foundamental} and \cite{String-Corrected}  are interested in finding the exact solutions of the truncated $\beta$ functions in a different content. However here  the truncated equations do not admit any exact solution with a perturbative expansion in $\alpha'$.} then this ambiguity would have allowed us to redefine the fields to  possibly obtain  an action  with either some divergence free covariant currents \cite{lovelock} or no exact unbounded linearized mode \cite{sen_foundamental}.  Note that ``T-duality symmetry is a well-defined notion over and above the presence of field redefinition ambiguity, in the sense that if it is present in one choice of scheme, it may be modified but will nonetheless also be present in any other scheme''\cite{Haagensen}. 

One way to obtain the $\alpha'$ corrections to T-duality is to require that T-duality must commute with the vanishing of the $\beta$ functions\cite{olsen,Schiapa}. However at the order $\alpha'^2$ this method requires lengthy  calculations. To avoid these calculations one can select different background  and calculate their  $\alpha'^2$ corrections and the $\alpha'^2$ corrections  to their corresponding T-dual backgrounds. Requiring T-duality to map these $\alpha'^2$ corrected background to each other enables one to write  the corrections to T-duality. In \cite{exir} the quadratic $\alpha'$ corrections to the general diagonal Kasner metric were computed and the  rules of T-duality in  $x_{25}$ for time-dependent backgrounds of diagonal metric and dilaton were found to be\footnote{Note that with this notation ``covariant'' derivative acts on the logarithm of the metric as it had been an scalar.}
\begin{eqnarray}\nonumber\\\label{ftda1}
\ln \tilde{g}_{_{25\,25}}&-&\frac{\alpha'}{4}\tilde{\nabla}_\mu\ln \tilde{g}_{_{25\,25}}.\tilde{\nabla}^\mu\ln \tilde{g}_{_{25\,25}}\\\nonumber\label{ftda2}
&=&-\left\{\ln {g}_{_{25\,25}}-\frac{\alpha'}{4}{\nabla}_\mu\ln {g}_{_{25\,25}}.{\nabla}^\mu\ln {g}_{_{25\,25}}\right\}+O(\alpha'^2\nabla^4)~,\\\nonumber\\
\ln \tilde{g}_{_{ii}}&+&\frac{\alpha'^2}{32}\tilde{\nabla}_\mu\ln \tilde{g}_{_{25\,25}}.\tilde{\nabla}^\mu\ln \tilde{g}_{_{ii}}~\tilde{\nabla}_\nu\ln \tilde{g}_{_{25\,25}}.\tilde{\nabla}^\nu\ln \tilde{g}_{_{25\,25}}\\
&=&\ln \tilde{g}_{_{ii}}+\frac{\alpha'^2}{32}{\nabla}_\mu\ln {g}_{_{ii}}.{\nabla}^\mu\ln {g}_{_{25\,25}}~{\nabla}_\nu\ln {g}_{_{25\,25}}.{\nabla}^\nu\ln {g}_{_{25\,25}}+O(\alpha'^3\nabla^6)\nonumber~,
\\\nonumber\\\label{ftda3}
\tilde{\phi}&-&\frac{1}{4}\ln \det\tilde g\,+\,\frac{\alpha'^2}{32}~\tilde{\nabla}_\mu\ln \tilde{g}_{_{25\,25}}~\tilde{\nabla}_\nu\ln \tilde{g}_{_{25\,25}}~\tilde{\nabla}^\mu\tilde{\nabla}^\nu\ln \tilde{g}_{_{25\,25}}\\
&=&\phi\,-\,\frac{1}{4}\ln \det g\,+\,\frac{\alpha'^2}{32}~\nabla_\mu\ln g_{_{25\,25}}~\nabla_\nu\ln g_{_{25\,25}}~\nabla^\mu\nabla^\nu\ln g_{_{25\,25}}~+~O(\alpha'^3\nabla^6)\nonumber~.
\end{eqnarray}
where ``$\tilde{g},\tilde{\phi}$'' and ``$g,\phi$''  represent the metric  and the dilaton  respectively of the background and its T-dual  in the co-moving frame in the string frame.  ُThe $\alpha'^2$ corrections to the Kasner metric are not sufficient to write uniquely the $\alpha'^2$ corrections to (\ref{ftda1}).  For  the diagonal Kasner metric in the string frame,
\begin{eqnarray}
ds^2  &=& - dt^2\,+\,\sum_{i=1}^{25}t^{2p_i}dx_i^2\,,\\
\phi(t) &=& \frac{\sum{p}-1}{2}\,\ln t\,,
\end{eqnarray}
 one notices that ``(\ref{ftda1})'' should  be modified to\cite{exir}
\begin{eqnarray}\label{tda12}
\ln \tilde{g}_{_{25\,25}}&-&\frac{\alpha'}{4}\,\tilde{\nabla}_\mu \ln \tilde{g}_{_{25\,25}}\tilde{\nabla}^\mu \ln\tilde{g}_{_{25\,25}}\,
\\
&=&-\left\{\ln g_{_{25\,25}}-\frac{\alpha'}{4}\nabla_\mu \ln g_{_{25\,25}}\nabla^\mu\ln g_{_{25\,25}}
+ \frac{\alpha'^2\,p_{_{25}}^4}{t^4}
-\frac{\alpha'^2{p}_{_{25}}^2}{t^4}\right\}\nonumber\,,
\end{eqnarray}
 There exists only one way to express $\frac{p_{_{25}}^4}{t^4}$ as the ``covariant'' derivative of the dilaton and the logarithm of the Kasner metric:
\begin{eqnarray}
\frac{p_{_{25}}^4}{t^4}&=&\frac{1}{16}(\nabla_\mu\ln g_{_{25\,25}}\,\nabla^\mu\ln g_{_{25\,25}})^2\,,
\end{eqnarray}
while there are the following five candidates for $\frac{p_{_{25}}^2}{t^4}$,
\begin{eqnarray}\label{exclude}
\label{A1}A^\star&=&\nabla_\mu \ln g_{_{25\,25}}\,\nabla^\mu \ln g_{_{25\,25}}~\nabla_\nu(\phi-\frac{1}{4}\ln\det g)\,\nabla^\nu(\phi-\frac{1}{4}\ln\det g)\,,\\
\label{A2}A&=&\nabla_\mu \ln g_{_{25\,25}}\,\nabla_\nu \ln g_{_{25\,25}}~\nabla^\mu(\phi-\frac{1}{4}\ln\det g)\,\nabla^\nu(\phi-\frac{1}{4}\ln\det g)\,,\\
\label{B}B&=&\frac{1}{2}\,\nabla_\mu\ln g_{_{25\,25}}\,\nabla_\nu\ln g_{_{25\,25}}\,\nabla^\nu\nabla^\mu(\phi-\frac{1}{4}\ln\det g)\,,\\
\label{C}C&=&\frac{1}{2}\,\nabla_\mu\ln g_{_{25\,25}}\,\nabla_\nu(\phi-\frac{1}{4}\ln\det g)\,\nabla^\nu\nabla^\mu\ln g_{_{25\,25}}\,,\\
\label{E}E&=&\frac{1}{8}\,\nabla_\mu\nabla_\nu\ln g_{_{25\,25}}\,\nabla^\mu\nabla^\nu\ln g_{_{25\,25}}\,.
\end{eqnarray}
In direct multiplication of spaces $|\nabla(\phi-\frac{1}{4}\ln\det g)|^2$ can depend on the coordinates of the individual spaces. We assume that T-duality rules for direct multiplications of curved spaces are given by the T-duality rules of each individual space. Therefore we exclude the possibility of writing $\frac{p_{_{25}}^2}{t^4}$ by (\ref{exclude}). Note that the rest of possibilities respect this assumption and  the general form of ``(\ref{tda12})'' must be a linear combination of them,  
\begin{eqnarray}\label{goal}
\ln \tilde{g}_{_{25\,25}}&-&\frac{\alpha'}{4}\,\tilde{\nabla}_\mu \ln \tilde{g}_{_{25\,25}}\tilde{\nabla}^\mu \ln\tilde{g}_{_{25\,25}}\,\\
&=&-\text{\huge\{}\ln g_{_{25\,25}}-\frac{\alpha'}{4}\nabla_\mu \ln g_{_{25\,25}}\nabla^\mu\ln g_{_{25\,25}}
+ \frac{\alpha'^2}{16}({\nabla}_\mu \ln {g}_{_{25\,25}}{\nabla}^\mu\ln{g}_{_{25\,25}})^2
\nonumber\\
&&-\,{\alpha'^2}\,(a\,{A}\,+\,b\,{B}\,+\,c\,{C}\,+\,e\,{E})\text{\huge\}}\,,\nonumber
\end{eqnarray}
where ``$a,b,c,e$'' are real numbers satisfying ``$a+b+c+e=1$''. The $\alpha'^2$ corrections to other backgrounds should be calculated to fix these numerical coefficients. The simplest background to be considered is a two-dimensional black hole, as a solution of the leading beta-functions  of the non-critical string theory :
\begin{eqnarray}\label{rmunu}
R_{\mu\nu}\,+\,2\,\nabla_\mu\nabla_\nu\phi&=&0\,,\\\label{alphaphi}
\frac{d-26}{6\alpha'}\,-\,\frac{1}{2}\,\Box\phi\,+\,|\nabla\phi|^2&=&0\,.
\end{eqnarray}
Using the convention``$\alpha'=\frac{26-d}{6}$'' in ``$d$'' dimensional space-time, this solution  reads\cite{Mueller,wadia}
\begin{eqnarray}\label{treelevel}
ds^2&=&dt^2\,+\,\tanh(t)^2\,dr^2\,+\,dx_{d-2}^2\,,\\
\phi(t)&=&-\ln(\cosh(t))\,,\nonumber
\end{eqnarray}
where  $dx_{d-2}$ is ``$d-2$'' flat directions. This solution with an appropriate periodicity ($r\sim r+2\pi$) is a fair candidate for a two-dimensional black hole with a geometry of semi-infinite cigar \cite{wadia,witten}.  We assume  that the non-critical string theory has the following perturbative\footnote{To make it perturbative, first we analytically extend the two-dimensional black hole to $d=26 - \epsilon$ where $\epsilon $ is a sufficiently small positive number. And at the end of the calculation we analytically extend the solution back to arbitrary $d$. } background  
\begin{eqnarray}
ds^2&=&dt^2\,+\,f(t)\,dr^2\,+\,dx_{d-2}^2\,,\\
f(t)&=&\tanh(t)^2\,(1\,+\,\alpha'\,f^{(1)}(t)\,+\,\alpha'^2\,f^{(2)}(t)\,+\,\cdots)\,,\\
\phi(t)&=&-\ln(\cosh(t))\,+\,\alpha'\,\phi^{(1)}(t)\,+\,\alpha'^2\,\phi^{(2)}(t)\,+\,\cdots\,.
\end{eqnarray} 
Inserting these perturbative series in the beta functions gives a set of ordinary linear differential equations for ``$f^{(1)}(t), f^{(2)}(t),\phi^{(1)}(t),\phi^{(2)}(t)$''.  These equations have unique answer for any given boundary condition. 
 Therefore the  assumptions do not contradict themselves in the sense that the set of differential equations for the $\alpha'$ terms is not overdetermined. We fix the boundary condition  assuming that there exists no correction at infinity,
 \begin{eqnarray}\label{boundary}
f^{(1)}(\infty)\,=\, f^{(2)}(\infty)\,=\,\phi^{(1)}(\infty)\,=\,\phi^{(2)}(\infty)\,=\,0\,.
\end{eqnarray} 
This boundary conditions results to the $\alpha'^2$ corrected background
 \begin{eqnarray}\label{fr}
f(t)&=&\tanh(t)^2\,(1\,-\,\frac{2\,\alpha'}{\cosh(t)^2}\,+\,\frac{\alpha'^2\,(5\,-\,\cosh(2t))}{\cosh(t)^4}\,+\,\cdots)\,,\\
\phi(t)&=&-\ln(\cosh(t))\,-\,\frac{\alpha'}{2\,\cosh(t)^2}\,-\,\frac{\alpha'^2\sinh(t)^2}{2\cosh(t)^4}\,+\,\cdots\,.\label{phir}
\end{eqnarray} 
 In the next section we  discuss in more detail the behaviour of the $\alpha'$ corrected metric.

 We are intersted in the modification of the T-duality so we consider the following background,
  \begin{eqnarray}
d\tilde{s}^2&=&dt^2\,+\,\frac{1}{\tanh(t)^2}\,dr^2\,+\,dx_{d-2}^2\,,\\
\tilde{\phi}(t)&=&-\,\ln(\sinh(t))\,,
\end{eqnarray}
which  is related to the two-dimensional black-hole by T-duality in the direction of $r$. Its  $\alpha'^2$ corrections follow
\begin{eqnarray}
d\tilde{s}^2 & =&dt^2\,+\,\tilde{f}(t)\,dr^2\,,\\
\tilde{f}(t)&=&\frac{1}{\tanh(t)^2}\,(1\,+\,\frac{2\alpha'}{\sinh(t)^2}\,+\,\frac{\alpha'^2\,(5\,+\,\cosh(2t))}{\sinh(t)^4})\,,\\
\tilde{\phi}(t)&=&-\,\ln(\sinh(t))\,+\,\frac{\alpha'}{2\,\sinh(t)^2}\,+\,\frac{\alpha'^2\,\cosh(t)^2}{2\,\sinh(t)^4}\,,
\end{eqnarray}
where it is assumed that the corrections vanish at infinity. The covariant form of T-duality rules  maps the  $\alpha'^2$ corrected two-dimensional black hole to its T-dual iff
\begin{eqnarray}\label{ambiguties}
a_2&=&-c\,,\\\nonumber
b&=&0\,,\\\nonumber
e&=&1\,.
\end{eqnarray}
 We set $a_2,b$ and $c$ to the above values. Furthermore the leading $\beta$ functions for  a time-dependent background\footnote{When the components of the metric depend only on time.} implies 
\begin{eqnarray}\label{identity}
c\,C\,+\,a_2\,A_2&=&c\,(C-A_2)\,\sim\,\beta_{_{25\,25}}\,=\,0\,.
\end{eqnarray}
Using ``(\ref{ambiguties})'' and ``(\ref{identity})'' simplifies the T-duality rule of the transverse direction  to\footnote{Also note that it is rewritten in the form that the square of the T-duality is the identity.}
\begin{eqnarray}\label{done}
\ln \tilde{g}_{_{25\,25}}&-&\frac{\alpha'}{4}\,\tilde{\nabla}_\mu\ln\tilde{g}_{_{25\,25}}\,\tilde{\nabla}^\mu\ln\tilde{g}_{_{25\,25}}
+\,\frac{\alpha'^2}{32}(\tilde{\nabla}_\mu\ln\tilde{g}_{_{25\,25}}\,\tilde{\nabla}^\mu\ln\tilde{g}_{_{25\,25}})^2\\
&-&\frac{\alpha'^2}{16}\tilde{\nabla}_\mu\tilde{\nabla}_\nu\ln\tilde{g}_{_{25\,25}}\,\tilde{\nabla}^\mu\tilde{\nabla}^\nu\ln\tilde{g}_{_{25\,25}}\nonumber\\
=&-&\left(\ln {g}_{_{25\,25}}\,-\,\frac{\alpha'}{4}\,{\nabla}_\mu\ln{g}_{_{25\,25}}\,{\nabla}^\mu\ln{g}_{_{25\,25}}
+\,\frac{\alpha'^2}{32}({\nabla}_\mu\ln{g}_{_{25\,25}}\,{\nabla}^\mu\ln{g}_{_{25\,25}})^2\right.\nonumber\\
&&~\left.-\,\frac{\alpha'^2}{16}{\nabla}_\mu{\nabla}_\nu\ln {g}_{_{25\,25}}\,{\nabla}^\mu{\nabla}^\nu\ln {g}_{_{25\,25}}\right)\,.\nonumber
\end{eqnarray}
This rule beside ``(\ref{ftda1})'' and ``(\ref{ftda2})'' provides the quadratic $\alpha'$ corrections to  T-duality on time-dependent backgrounds of  diagonal metric and dilaton. As a check of the consistency of our procedure these rules  have been checked to be true for the Schwarzschild metric in ``$D=4,5$''  dimensions and T-duality in the direction of the time-like Killing vector outside the horizon\footnote{The  $\alpha'^2$ corrections to Schwarzschild black hole and its T-dual in $D=5$ are presented in the appendix,  for $D=4$ the results can be found in \cite{exir}}. It is not necessary to check these rules on different backgrounds since the explicit computation of \cite{odd} supports the expansion of T-duality rules at order $\alpha'^2$.

At this stage it is natural  to ask if the $\alpha'$ corrections to T-duality can be canceled by an appropriate field redefinition. In other words, is there any appropriate renormalization scheme and regularization method which gives no correction to the tree level rules of T-duality? The answer is negative. One can check that  even the linear $\alpha'$  corrections to T-duality can not be compensated by a field redefinition which leaves the tensor property of the metric intact. In the former works \cite{oddtwoloop,odd} either the redefined metric had not been a tensor or the metric definitions had not been the same  in both spaces. Choosing different definitions for the metrics means choosing different schemes for the background and for its T-dual. If we do so then the corrections in the space can not be directly mapped to the corrections in the T-dual space. Choosing a scheme in which the metric is not a tensor means choosing a regularization method which breaks the general covariance of the theory. In such a regularization method one must be extremely careful about interpreting the results.  It is preferable to work in a scheme which respects the fundamental symmetries of the theory. In such a  scheme T-duality must be modified. This argument is supported  by the fact that T-duality rules are modified for the conjectured $\alpha'$ exact backgrounds\cite{verlinde}.

\section{The $\alpha'^2$ corrected two-dimensional metric} 
It is known that the string coupling constant is related to the dilaton. Hence the string coupling constant receives $\alpha'$ corrections through the corrections to the dilaton,
\begin{eqnarray}
g_s&=&g_{_0}\,e^{\phi+\alpha'\phi^{(1)}+\alpha'^2\phi^{(2)}+\cdots}\,.
\end{eqnarray}
Thus a small tree level string coupling constant generally does not guarantee a small string coupling constant. In the case of the two dimensional black hole one sees that  the $\alpha'$ corrections to the string coupling constant are finite. Therefore  the string coupling constant  will be small if the tree level one is small. Hence the string loop effects can be consistently ignored not only at the tree level\cite{onformation} but also to, at least, the quadratic order in $\alpha'$.
 
The   $\alpha'$ and $\alpha'^2$ corrected metrics become zero at some positive values of $t$ near ``$t=0$''. These zeros correspond to intrinsic singularities since their corresponding Ricci scalars diverge.   A similar behavior has been reported in \cite{finiteT} where  AdS/CFT correspondence is used to calculate specific quantum corrections to the general smooth half BPS solutions of type IIB supergravity\cite{LLM} and it is observed that the quantum corrections can change the coordinate singularity to either an intrinsic singularity or a highly curved geometry.

However the singularities of the two-dimensional black hole occur at the region where the tree level curvature is of the order of the string length. Thus no concrete conclusion can be achieved unless one  takes into account all the $\alpha'$ corrections. It should be useful to generalize \cite{borninfledgravity,Grumiller} toward the analog of the DBI action \cite{BDF} for gravity coupled to dilaton. In the perturbative studies if one insists on having a smooth $\alpha'$ corrected two-dimensional black hole  then one can choose either of the following methods,
\begin{enumerate}
\item  The $\alpha'$ corrected metric could be smooth in a different coordinate system. For example one can write the metric in the following form
\begin{eqnarray}
ds^2&=&(1+\frac{2\alpha'}{\cosh(t)^2}+\frac{2\sinh(t)^2\alpha'^2}{\cosh(t)^4}+O(\alpha'^3))\,dt^2\,+\,\tanh(t)^2\,dr^2\,+\,dx_{d-2}^2\,,\end{eqnarray}
In this new set of coordinates one does not face any singularity if one extrapolates the $\alpha'$ metric from infinity toward the origin.

\item One can choose the  boundary conditions requiring a  smooth geometry . A boundary condition which is different from (\ref{boundary}) results in
\begin{eqnarray}
ds^2&=&dt^2+\tanh(t)^2(1+\alpha'C_1+\alpha'^2C_2-\frac{2\alpha'}{\cosh(t)^2}-\frac{\alpha'^2(5-\cosh(t))}{\cosh(t)^4})dr^2\nonumber\,,\nonumber
\end{eqnarray}
where $C_1$ and $C_2$ are arbitrary real numbers. These numbers  can be chosen in such a way that the metric remains nonzero. This choice alters the metric at infinity.

\item One can employ the field redefinition ambiguity to obtain a regular metric. For example redefining the metric by
\begin{eqnarray}
g_{\mu\nu}&\to&g_{\mu\nu}\,+\,\alpha'\,R_{\mu\nu}\,+\,\alpha'^2\,(\frac{1}{2}R_{\mu\eta}R^{\eta}_{~\nu}+\frac{3}{8}R\,R_{\mu\nu}-\frac{1}{4}\Box R_{\mu\nu}+\frac{3}{8}\nabla_\mu\nabla_\nu R)\nonumber\,,
\end{eqnarray}
gives
\begin{eqnarray}
ds^2&=&(1+\frac{2\alpha'}{\cosh(t)^2}+\frac{4\alpha'^2}{\cosh(t)^2})dt^2\,+\,\tanh(t)^2\,dr^2\,,
\end{eqnarray}
where it is clear that the geometry of the redefined metric is smooth. One may ask if the $\beta$ functions of  such a redefined metric  have   better properties. For example, do all the exact modes of the linearized action have real frequency \cite{sen_foundamental}?, or do the beta functions  admit  covariant divergence free currents\cite{lovelock}? These questions need further investigation.  

\item We can rewrite the $\alpha'$ expansion series in such a way that the components of the metric remain nonzero. For example \cite{foundattheend} uses a different series  which gives a regular metric. Note that as shown in \cite{foundattheend} the linear $\alpha'$ corrections are the same as perturbative expansion of \cite{verlinde} in which a modification of the two-dimensional black hole is conjectured to be $\alpha'$-exact while the agreement at the quadratic order requires a field redefinition.
\end{enumerate}
Also a combination of the above methods can be chosen to obtain a regular $\alpha'$ corrected background.

\section{Conclusions and discussions}
In the second section we computed the quadratic $\alpha'$ corrections to the two-dimensional black hole and to its T-dual. These $\alpha'^2$ corrected backgrounds and the $\alpha'^2$ corrected Kasner metric\cite{exir} were used to write the covariant form of the $\alpha'^2$ corrections to T-duality for general time-dependent background of diagonal metric and dilaton in bosonic string theory. It was shown that  no appropriate  perturbative field redefinition compensates the corrections to T-duality. Thus tree level rules of T-duality must be modified. This argument was supported  by the fact that T-duality rules are modified for the conjectured $\alpha'$ exact backgrounds\cite{verlinde}. 

In the third section we showed that the $\alpha'$ correction to the string coupling constant is finite for the two-dimensional black hole. Therefore the string loop effects  can be ignored  consistently not only at the tree level \cite{onformation} but also to, at least, the quadratic order in $\alpha'$. It was observed that the $\alpha'$ corrections may change the horizon to an intrinsic singularity. However no concrete conclusion was achieved due to the fact that  the curvature at the horizon is comparable to the string length and one must take into account all the $\alpha'$ corrections. Note that the perfect reflection of the tachyon wave \cite{onformation} points to the possibility of the absence of a horizon, while the conjectured $\alpha'$ exact background\cite{verlinde} supports the existence of the horizon.

We presented four different methods to preserve the horizon in the perturbation theory. In addition to these possibilities other fields may be required to be turned on  besides the field content of the  tree level background (\ref{treelevel}). To clarify this statement  recall that the subleading corrections to the $\beta$ function of the dilaton  can induce a non-trivial dilaton at the subleading level when  the tree level dilaton vanishes. A priori it is not clear if all other modes, in particular the massless tachyon, can be consistently set to zero to all orders in $\alpha'$. In the context of WZW coset models, the  conjectured $\alpha'$-exact  two-dimensional black hole \cite{verlinde}  has only non trivial  dilaton and metric. However in the context of MQM ref. \cite{Vladimir} argues that   certain winding modes of Tachyon should be turned on for the two-dimensional black hole, this possibility is also emphasized in   \cite{onformation,ads}.

It is a hopeless task to calculate the $\alpha'$ perturbative expansion of the beta functions of a general background, a background which includes  all the excitation of string theory. It is easier to guess the field contents of the underlying conformal theory than to ask if the corresponding beta functions have a perturbative solution. When the answer is affirmative, like the case of the two-dimensional black hole, the guess can be correct.

\section*{Acknowledgments}
I thank Loriano Bonora and Martin O'loughlin for both the suggestions and the discussions throughout the work, and  Carlo Maccaferri for discussions. I thank Ricardo Schiappa and Daniel Grumiller  for the correspondences. 

 \appendix

\section{The $\alpha'^2$ corrections to the Schwarzschild metric in $D=5$}
The five dimensional Schwarzschild black hole  reads
\begin{equation}
ds^2\,=\,-(1-\frac{1}{r^2})\,dt^2\,+\,\frac{dr^2}{1-\frac{1}{r^2}}\,+\,r^2\,d\Omega_3\,,
\end{equation}
where the mass of the black hole is appropriately chosen to give a factor of one in $\frac{1}{r^2}$. For very large $r$ this metric can be generalized to a perturbative background of free critical bosonic string theory,\footnote{The linear $\alpha'$ corrections to the four dimensional black hole were computed in \cite{callan}.}
\begin{eqnarray}
ds^2&=&-\,f(r)\,dt^2\,+\,g(r)\,dr^2\,+\,r^2\,d\Omega_3\,+\,{\overrightarrow{dx}_{_{21}}}^{2}\,,\\
f(r)&=&(1-\frac{1}{r^2})\,(1\,+\,\alpha'\,f^{(1)}(r)\,+\,\alpha'^2\,f^{(2)}(r)\,+\,\cdots)\,,\\
g(r)&=&\frac{1}{1-\frac{1}{r^2}}(1\,+\,\alpha'\,g^{(1)}(r)\,+\,\alpha'^2\,g^{(2)}(r)\,+\,\cdots)\label{gr}\,,\\
\phi(r)&=&0\,+\,\alpha'\,\phi^{(1)}(r)\,+\,\alpha'^2\,\phi^{(2)}(r)\,+\,\cdots\,,
\end{eqnarray}
where $\overrightarrow{dx}_{_{21}}$ represents the 21 flat directions compactified to a torus and $\phi(r)$ is the dilaton. Using the beta functions and assuming that there is no correction to the mass of the black hole or to the fall off of the dilaton at infinity, one finds
\begin{eqnarray}
\phi^{(1)}(r)&=&\frac{25}{8}\ln(\frac{r^2\,+\,1}{r^2\,-\,1})\,-\,\frac{25}{12\,r^6}\,-\,\frac{25}{4\,r^2}\,,\\
f^{(1)}(r)&=&\frac{1}{6\,r^6\,(r^4-1)}\left(48+80r^4-30 r^8 + 15 r^6 (r^4-3) \ln(\frac{r^2+1}{r^2-1})\right)\,,\\
g^{(1)}(r)&=&\frac{1}{3 r^6 (r^4-1)}\left(11 - 30 r^4 + 15 \ln(\frac{r^2 + 1}{ r^2-1 })\right)\,,
\end{eqnarray} 
and
\begin{eqnarray}
\phi^{(2)}(r)&=&\frac{2364 r^8 - 2478 r^6 - 394 r^4 - 953 r^2 - 915}{768\, r^8\, (r^2-1)}\,+\,\\
&&\,+\,\frac{\ln(1-\frac{1}{r^2})}{64 r^4 (r^2-1)}(197 r^6 - 305 r^4 + 54 r^2 -54)\nonumber\,,\\
f^{(2)}(r)&=& \frac{ 1068\,r^{10}\,-\,3966 r^{8}\,+\,1214\,r^{6}\,-\,435\,r^{4}\,-\,1553 r^{2}\,+\,6048}
{576\,(r^{2}\, -\, 1)\, r^{10}}\,+\,\\
&&\,+ \,\frac {\ln
(1 - { \frac {1}{r^{2}}} )}{48\,r^{12}\,(r^{2} - 1)^{2}}\,(89\,r^{8} - 321\,r^{6}
 + 16\,r^{4} - 108\,r^{2} + 432)\,+\,  \nonumber\\
&&+\frac{9\,(r^2\,-\,4)}{8}(\ln(1-\frac{1}{r^2}))^2\nonumber \,,\\
g^{(2)}(r)&=&{\displaystyle \frac {876\,r^{8} + 210\,r^{6}
 + 574\,r^{4} + 6449\,r^{2} - 4473}{576\,r^{6}\,(r^{2} - 1)^{3}}
}\,+\,\\
&&\,-\,  {\displaystyle \frac {(73\,r^{4} - 73\,r^{2} + 144)}{48\,(r^{2} - 1)^{3}}}\,\ln(1 -  \frac {1}{r^{2}} )\,+ \,{ \frac {9\,r^2\,(r^{2} + 1)}{8\,(r^{2} - 1)^{3}}}\,\ln(1 - \frac {1}{r^2} )^{2} \nonumber\,.
\end{eqnarray}
Note that the $\alpha'$ corrected metric has a singularity outside the horizon.  This is reminiscent of what happens in the case of two-dimensional black hole. In addition the $\alpha'$ corrected string coupling constant diverges at $r=1$. We come back to these issues at the end of this appendix. Now let us apply
 T-duality in the direction of the time-like Killing vector  outside the horizon 
\begin{eqnarray}
d\tilde{s}^2&=&-\,\frac{dt^2}{1-\frac{1}{r^2}}\,+\,\frac{dr^2}{1-\frac{1}{r^2}}\,+\,r^2\,d\Omega_3\,,\\
\tilde{\phi}(r)&=&0\,.
\end{eqnarray}
The T-dual of the Schwarzschild metric is singular at ``$r=1$'' which means that the coordinate singularity of the Schwarzschild metric has changed to an intrinsic singularity. Since T-duality relates the singular time-dual  geometry to the Schwarzschild metric then  one expects that the time-dual metric to be as stable as the Schwarzschild metric \cite{stability}. This  argument should not sound strange because the classical stability of a naked singularity   recently has been explored  in \cite{gibbons}.  At large $r$ T-dual background can be generalized to a perturbative background of string theory whose quadratic $\alpha'$ corrections follow \footnote{We have chosen this specific coordinate since it is easier to apply T-duality in this coordinate.}
\begin{eqnarray}
d\tilde{s}^2&=&-\,\tilde{f}(r)\,dt^2\,+\,g(r)\,dr^2\,+\,r^2\,(1+\frac{\alpha'^2}{r^8\,(r^2\,-\,1)})\,d\Omega_3\,+\,{\overrightarrow{dx}_{_{21}}}^{2}\,,\\
\tilde{f}(r)&=&\frac{1}{1-\frac{1}{r^2}}\,(1\,+\,\alpha'\,f^{(1)}(r)\,+\,\alpha'^2\,f^{(2)}(r)\,+\,\cdots)\,,\\
\tilde{\phi}(r)&=&0\,+\,\alpha'\,\tilde{\phi}^{(1)}(r)\,+\,\alpha'^2\,\tilde{\phi}^{(2)}(r)\,+\,\cdots\,,
\end{eqnarray}
where we are going to assume that there is no $\alpha'$-correction to the fall off of $f(r)$ at infinity. Note that $g(r)$ is provided by (\ref{gr}). The linear and the quadratic terms in $\alpha'$ in $\tilde{f}(r)$ follow
\begin{eqnarray}
\tilde{f}^{(1)}(r)&=&{\frac{6\,r^2\,-\,9}{4\,(r^2\,-\,1)\,r^2}\,+\,\frac{3\,(r^2\,-\,2)}{2\,(r^2\,-\,1)}}{\,\ln(1-\frac{1}{r^2})}\,,\\
\tilde{f}^{(2)}(r)&=&-\frac{1068\,r^{10}\,-\,5262\,r^8\,+\,5102\,r^6\,-\,1623\,r^4\,+\,1039\,r^2\,+\,576}{576\,r^8\,(r^2\,-\,1)^2}\,+\\
&&\,-\,\frac{89\,r^4\,-\,448\,r^2\,+324}{48\,r^2\,(r^2\,-\,1)}\ln(1-\frac{1}{r^2})\,+\,\frac{9(r^4 -3 r^2 +4)}{8\,(r^2-1)^2}(\ln(1-\frac{1}{r^2}))^2\nonumber\,.
\end{eqnarray}  
And the coefficients of $\alpha'$ in $\tilde{\phi}(r)$ read
\begin{eqnarray}
\tilde{\phi}^{(1)}(r)&=&-\frac{6\,r^4\,+\,9\,r^2\,-1}{16\,r^4\,(r^2\,-\,1)}-\frac{3\,(r^2\,+\,1)}{8\,(r^2\,-\,1)}\ln(1-\frac{1}{r^2})\,,\\
\tilde{\phi}^{(2)}(r)&=&\frac{4956 r^{10} -5298 r^8 -64 r^6 - 1347 r^4 + 1492 r^2 - 2151}{2304 (r^2-1)^2 r^8}\,+\,\\
&&\,+\frac{413 r^8-648r^6+289r^4-36r^2+18}{194\,r^4\,(r^2\,-\,1)}\ln(1-\frac{1}{r^2})\,
+\nonumber\\&&\,
+\frac{9\,r^2}{16\,(r^2\,-\,1)^2}(\ln(1-\frac{1}{r^2}))^2\nonumber\,.
\end{eqnarray}
Instead of writing the $\alpha'$ corrected Schwarzschild black hole and its T-dual metric  in the ``radial" co-moving frame, it is easier to rewrite the rules of   T-duality for metrics which can be transform to the ``radial" co-moving frame by the same coordinate transformation. Here the $\alpha'$ corrected Schwarzschild black hole and its T-dual can be transformed to the ``radial'' co-moving frame by 
 the same coordinate transformation. Luckily this transformation does not alter the rules of T-duality. It is a straightforward computation to check that the $\alpha'$ corrected T-duality rules (\ref{ftda1},\ref{ftda2},\ref{done}) relate the $\alpha'$ corrected black-hole to its T-dual background.  
 
The $\alpha'$ corrected metric is singular outside the horizon.  If one insists on having a smooth geometry then one can choose the boundary conditions by requiring finite corrections  at $r=1$,
\begin{eqnarray}\label{lastname}
ds^2&=&-f^{\star}(r)\,(1-\frac{1}{r^2})\,dt^2\,+\,\frac{g^{\star}(r)}{1-\frac{1}{r^2}}\,dr^2\,+\,r^2\,d\Omega^2_3\,,\\
f^{\star}(r)&=&1+c_1\alpha'+c_2\alpha'^2-\alpha'\frac{17r^2+8}{4r^4}-\alpha'^2  {\displaystyle \frac {1039\,r^{6} - 4811\,r
^{4} - 10543\,r^{2} - 6048}{576\,r^{8}}}\,,\nonumber\\
g^{\star}(r)&=&1-\alpha'\, {\displaystyle \frac {r^{2} + 7}{4\,r^{4}}} \,+\,\alpha'^2\,{\displaystyle \frac {355\,r^{6} - 1565\,r^{4}
 + 2497\,r^{2} + 4473}{576\,r^{8}}} \,,\nonumber\\
\phi^{\star}(r)&=&0\,-\,\alpha'  {\displaystyle \frac {9\,(1 + 2\,r^{2})}{16\,r^{4
}}}    \,-\,\alpha'^2  {\displaystyle \frac {228\,r^{6} + 6\,r^{4} - 
1868\,r^{2} - 915}{768\,r^{8}}}\,,\nonumber
\end{eqnarray}
where $c_1$ and $c_2$ are numerical constants. These boundary conditions also gives finite corrections to the string coupling constant. One may fix ``$c_1=c_2=0$''  assuming no correction exist  at infinity. The price of having a smooth geometry is to change the fall off of the metric at infinity. The asymptotic behaviour of the linear $\alpha'$ corrected metric at large $r$ in the Einstein frame is
\begin{eqnarray}
ds_E^{\star2}&=&(1+2c_1\alpha'-\frac{1+\frac{11}{4}\alpha'}{r^2})\,dt^2\,+\,(1+\frac{1+\frac{5}{4}\alpha'}{r^2})dr^2\,+\,r^2\,d\Omega^2_3\,+\,O(\alpha'^2)\,+\,O(\frac{1}{r^4})\nonumber\,,
\end{eqnarray}
The corrections to the T-dual background are
\begin{eqnarray}
d\tilde{s}^{\star2}&=&-\frac{\tilde{f}^{\star}(r)}{1-\frac{1}{r^2}}\,dt^2\,+\,\frac{g^{\star}(r)}{1-\frac{1}{r^2}}\,dr^2\,+\,r^2(1\,+\,\frac{\alpha'^2}{r^8\,(r^2-1)})\,d\Omega^2_3\,,\\
\tilde{f}^{\star}(r)&=&1\,-\,\alpha' \frac {17\,r^{2} - 9}{4\,r^{2}\,(r^{2} - 1)} 
+\nonumber\\
&&-\,\alpha'^2 \frac { - 576 + 1623\,r^{4} - 1039\,r^{2} + 3515\,r^{8} - 1106\,r^{6} + 1039\,r^{10}}{576\,r^{8}\,(r^2 - 1)^{2}} \nonumber\,,\\
\tilde{\phi}^{\star}(r)\,&=&\,- {\displaystyle \frac {1}{2}} \,\mathrm{ln}(1 - {\displaystyle \frac {1}{r^{2}}} )\,+
\,\alpha'{\displaystyle \frac {16\,r^{4} - 9\,r^{2} + 1}{16\,r^{4}\,(r^{2} - 1)}} \,+\nonumber\\
&&+\,\alpha'^2\,{\displaystyle \frac { - 1527\,r^{4} + 1492\,r^{2}
 - 2024\,r^{8} + 3968\,r^{6} + 1394\,r^{10} - 2151}{2304\,r^{8}\,
(r^2 - 1)^{2}\,}}\nonumber\,.
\end{eqnarray} 
 These expressions are consistent with $\alpha'$ corrected T-duality. We could have used T-duality  to find the corrections, by the same method that \cite{confusing} finds the linear $\alpha'$ corrections to the T-dual of the four dimensional black hole.

\end{document}